# Temperature dependence of electrical and thermal conduction in single silver nanowire

Zhe Cheng[1], Longju Liu[2], Meng Lu[2], Xinwei Wang[1]*

[1]Department of Mechanical Engineering, 2010 Black Engineering Building,

Iowa State University, Ames, Iowa, 50011, USA

[2]Department of Electrical and Computer Engineering, Iowa State University, Ames, Iowa, 50011,

USA

**Abstract**

Silver nanowires have great application potential in fields like flexible electronic devices, solar cells and transparent electrodes. It is critical and fundamental to study the thermal and electrical transport properties in a single silver nanowire. In this work, the thermal and electrical transport in an individual silver nanowire is characterized down to 35 K with the steady state electro-thermal technique. The results indicate that, at room temperature, the electrical resistivity increases by around 4 folds compared from that of its bulk counterpart. After fitting the temperature dependent electrical resistivity curves with the Bloch-Grüneisen formula, the Debye temperature (151 K) of the silver nanowire is found 36% lower than that (235 K) of bulk silver, confirming strong phonon softening. The thermal conductivity is reduced by 55% compared with that of its bulk counterpart at room temperature and this reduction becomes larger as the temperature goes down. To explain the opposite trends of thermal conductivity ($\kappa$) ~ temperature ($T$) of silver nanowire and bulk silver, a unified thermal resistivity ($\Theta = T/\kappa$) is used to elucidate

---

*Corresponding author. Email: xwang3@iastate.edu, Tel: 515-294-2085, Fax: 515-294-3261

the electron scattering mechanism. A large residual unified thermal resistivity for the silver nanowire is observed while that of the bulk silver is almost zero. The same trend of $\Theta$ variation against $T$ indicates that the silver nanowire and bulk silver share the same phonon-electron scattering mechanism. Additionally, due to phonon-assisted electron energy transfer across the grain boundaries, the Lorenz number of the silver nanowire is found much larger than that of bulk silver and decreases with decreasing temperature.

**Keywords**: silver nanowire; electrical transport; thermal conductivity; Lorenz number; unified thermal resistivity; electron scatterings2the electron scattering mechanism. A large residual unified thermal resistivity for the silver nanowire is observed while that of the bulk silver is almost zero. The same trend of $\Theta$ variation against $T$ indicates that the silver nanowire and bulk silver share the same phonon-electron scattering mechanism. Additionally, due to phonon-assisted electron energy transfer across the grain boundaries, the Lorenz number of the silver nanowire is found much larger than that of bulk silver and decreases with decreasing temperature.

# 1. Introduction

Recently silver nanowire has attracted considerable attention due to its great potentials for applications like flexible touch screen, solar cells and transparent electrodes.[1-5] For the design and optimization of these applications, the thermal and electrical properties of an individual silver nanowire are critical and fundamental but they have been rarely reported. Up to now, the electrical properties of nanowires, especially inert metallic nanowires and nanowire bundles, are not difficult to measure.[6-9] But for thermal property characterization, only a few experimental investigations have been reported due to the difficulties in suspending a single nanowire, reducing contact resistance and accurate thermal measurement. Ou *et al.* investigated the thermal and electrical conductivities of a single nickel nanowire from 15 K to 300 K. Its Lorenz number is larger than Sommerfeld value with temperature above 75 K and decreased rapidly when the temperature goes below 75 K.[10] The experimental results of Völklein *et al.* showed the thermal and electrical conductivities of a single Pt nanowire from 260 K to 360 K. The Lorenz number was smaller than the Sommerfeld value while the reasons are unclear in their work. [11] Stojanovic *et al.* measured the thermal conductivity of aluminum nanowire arrays instead of a single nanowire at room temperature. The thermal conductivity of these nanowire arrays is measured as 105-145 W/m·K when the widths of the nanowires ranged from 75 nm to 150 nm. The phonon contribution to the total thermal conductivity is about 21 W/m·K.[12]

For metallic nanostructures, when the system size approaches or is smaller than the bulk electron mean free path, the grain boundary and surface become important scattering sources which limit the electron mean free path and subsequently decrease the thermal and electrical conductivities. The scattered electrons cannot pass through the grain boundary but they can exchange energy with local phonons via electron-phonon scattering. Phonons can transport through the grain boundary more readily than electrons. Consequently, part of the scattered electrons' energy is transferred across the grain boundary via phonons while the charge of the scattered electrons is reflected back. This leads to the significantly reduced electrical conductivity and less reduced thermal conductivity. The resulting Lorenz numbers of metallic nanostructures are larger than the Sommerfeld value. The large Lorenz numbers are observed in most experimental investigations.[10, 13-18] But bulk-like Lorenz numbers are also observed experimentally for metallic nanofilms deposited on bio-materials like milkweed fiber and silkworm silk. This is explained by the electron tunneling and hopping via the saturated molecules or/and conjugated molecules in these bio-material substrates. [19-21]

In this work, the thermal and electrical properties of a single suspended silver nanowire are characterized with a steady state electro-thermal technique from room temperature down to 35 K. The temperature dependent Lorenz number is also determined. The thermal and electrical conductivities of the nanowire are compared with their bulk counterpart and a unified thermal resistivity is used to reveal the scattering mechanism.

**2. Sample preparation and structure**

Silver nanowires used in this work were purchased from Sigma-aldrich and they were supplied as suspensions in isopropyl alcohol (IPA) with a concentration of 0.5%. The purchased dispersion was further diluted with IPA and dropped on a piece of gel film. The silver nanowires remained on the gel film after IPA evaporated. In this process, the purchased dispersion should be diluted to a degree that single silver nanowires stayed on the gel film without contact with the surrounding ones. Then we used a very simple but effective way to make electrodes for suspending a single nanowire. First of all, a 180 nm thick Au film was deposited on a thermal oxide silicon wafer with a 1 μm thick $SiO_2$ layer. The wafer ($Au/SiO_2/Si$) was cut into pieces and assembled into two electrodes on a glass slide. The trench width is adjustable and the minimum trench width by this method is 5 μm. If the width is smaller than 5 μm, the two electrodes would be prone to connect with each other electrically. The width used in this work was about 25 μm. The two ends of the trench were fixed to the glass slide by epoxy glue.

Before bridging the nanowire, the two electrodes were checked to make sure that they did not connect with each other electrically. After that, a probe station was used to manipulate a single silver nanowire and bridge it across the electrodes. Finally, to suppress thermal and electrical contact resistance, Electron Beam Induced Deposition (EBID) was used to deposit Pt on the two ends of the silver nanowire. To guarantee the good thermal and electrical contact between the silver nanowire and the electrodes, the Pt pads are large and thick enough to cover the entire silver nanowire ends. For the thermal and electrical characterization of the single silver nanowire at low temperatures, a cryogenic system [CCS-450, JANIS] was used to provide cryogenic experiment environment as low as 10 K. The sample was put in a vacuum chamber to suppress the convection heat transfer. The pressure of the chamber is below 0.5 mTorr. The schematic diagram and SEM picture of the two electrodes and the silver nanowire sample (top view) are shown in Fig. 1(a) and (b). The side views are shown in Fig. 1(c) and (d). According to the top view and side view of the silver nanowire, the three dimension coordinates of the nanowire can be extracted and the length of the suspended silver nanowire is calculated as 27.23 μm. Unlike the length, the average nanowire diameter is easy to measure using SEM and its value is determined as 227 nm.

To characterize the structure of the silver nanowire, x-ray diffraction (XRD) was used to scan the sample. The XRD system (Siemens D 500 diffractometer) is equipped with a copper tube that was operated at 40 kV and 30 mA. Because a single silver nanowire is too small compared with the XRD spot size and cannot provide enough signals, five drops of purchased dispersion which contained plenty of silver nanowires were dropped on the XRD sample holder. The XRD pattern is shown in Fig. 2. The XRD pattern shows that the silver nanowire is composed of crystals. According to the pattern, the crystal size can be estimated based on the peaks. The crystal size calculated from Peak (111) is 126 nm and those sizes calculated from Peak (220) and Peak (311) are 8 nm and 21 nm respectively. All of the crystal sizes in these directions are much smaller than the diameter of the silver nanowire. This proves that the silver nanowire is polycrystalline. The grain boundaries among these nanocrystals in the silver nanowires are abundant. Also the different crystallite size determined by the peaks indicates that the crystallite in the nanowire is not cube-like or sphere-like. Instead, the crystallite is expected to be ellipsoid-like.

## 3. Electrical conduction

For the electrical characterization of the silver nanowire, a set of small electrical currents ranging from 0.1 mA to 0.5 mA were applied. The current source is Keithley 6221 DC and AC Current Source. Due to the electrical heating effect, the electrical resistance would rise when the electrical current increases. The measured electrical resistance should change proportionally with the electrical current's square. A linear fitting was used to extrapolate the electrical resistance without heating effect, namely the resistance when the electrical current is zero. More details can be found in the thermal characterization part which will be discussed later. The temperature dependent electrical resistance and electrical resistivity are shown in the inset of Fig. 3 and the electrical resistivity is fitted with the Bloch-Grüneisen formula. To compare the electrical resistivity of the silver nanowire and that of the bulk silver, both of them are shown in Fig. 3 and fitted with the Bloch-Grüneisen formula. [22] According to the Matthiessen's rule and the Bloch-Grüneisen theory,[23] the electrical resistivity can be expressed as $\rho=\rho_0+\rho_{el-ph}$. $\rho_0$ is the residual resistivity which results from structural scatterings, like grain boundary scattering, impurity scattering and surface scattering. It is essentially temperature independent and its value is the electrical resistivity when the temperature approaches zero. $\rho_{el-ph}$ is the temperature dependent electrical resistivity induced by phonon scattering, and can be expressed as

$$\rho_{el-ph} = \alpha_{el-ph}\left(\frac{T}{\theta}\right)^n \int_0^{\theta/T} \frac{x^n}{(e^x-1)(1-e^{-x})}dx, \tag{1}$$

where $\alpha_{el-ph}$ is the electron-phonon coupling parameter. $\theta$ is the Debye temperature and $n$ generally takes the value of 5 for nonmagnetic metals.[8] Through fitting the experimental data, the Debye temperature and the electron-phonon coupling constant can be obtained respectively. The Debye temperature of the silver nanowire and the bulk silver is 151 K and 235 K and the electron-phonon coupling constant of the silver nanowire and the bulk silver is $9.90\times10^{-8}$ Ω·m and $5.24\times10^{-8}$ Ω·m. The excellent fitting curves indicate that the phonon-electron scattering dominates the temperature dependent part of the electrical resistivity.

We can see from Fig. 3 that the residual electrical resistivity of the bulk silver is almost zero ($1\times10^{-11}$ Ω·m) while that of the silver nanowire is much larger ($3.25\times10^{-8}$ Ω·m). The electrical resistivity of silver nanowire at low temperatures is more than three orders of magnitude larger than that of the bulk silver. This is due to the intensive structural electron scatterings like grain boundary scattering and surface scattering. Because at low temperatures the phonon scattering diminishes, only structural scatterings contribute to impede electron transport. The electrical resistivity of silver nanowire at room temperature is five times as large as the counterpart of the bulk silver. This is due to the combined effect of different structural and phonon scatterings. For bulk silver, the phonon-electron scattering dominates at room temperature. But for the silver nanowire, both the phonon scattering and structural scatterings contribute to the large electrical resistivity.

It can be seen from the inset of Fig. 3 that the electrical resistivity changes linearly with temperature when temperature is not very low. The slope of the silver nanowire's electrical resistivity against temperature ($1.68 \times 10^{-10}$ Ω·m/K) is much larger than that of the bulk silver ($6.11 \times 10^{-11}$ Ω·m/K). Similar phenomenon is also observed in Co/Ni Superlattices.[24] Furthermore, the electrical resistivity of the silver nanowire and the bulk silver are both fitted well with the Bloch-Grüneisen theory. The fitting results show that the Debye temperature of the silver nanowire (151 K) is much smaller than that of the bulk silver (235 K). The reduced Debye temperature is due to surface phonon softening. The missing bonds of atoms at the surfaces, including inner surfaces like grain boundaries, lead to the change of phonon modes and vibration frequency. These changes result in the reduced Debye temperature. [6, 7, 24-26]

The electrical conductivity of silver is $\rho = m\tau^{-1}/(ne^2)$. Here, $m$ and $e$ is the electron mass and charge; $\tau$ is the relaxation time and $n$ is the electron density. When the temperature approaches zero, the effect of phonon scattering would diminish and the structural scatterings dominate in electron transport. The residual resistivity can be written as $\rho_0 = m\tau_0^{-1}/(ne^2)$. The electron density of silver is $5.85 \times 10^{28}$ m$^{-3}$.[27] The relaxation time is $\tau_0 = 1.87 \times 10^{-14}$ s. The Fermi velocity of silver is $1.39 \times 10^6$ m/s. [27] So the electron mean free path induced by the structural scatterings based on the residual electrical resistivity is 26 nm. This characterization length is close to the crystal size (21 nm) in the direction (311). The electron transport direction in our work is along the axial direction of the silver nanowire, we can conclude that axial direction is along (311).

## 4. Temperature dependence of thermal conductivity

*4.1 Thermal characterization method*

The thermal conductivity of the nanowire is characterized by using the steady-state electro-thermal technique. The silver nanowire is suspended across the two electrodes in the characterization. We can consider the two electrodes as two heat sinks. Its temperature is the same as the environment temperature $T_0$. The sample is placed in a high vacuum chamber to suppress the convection effect. Moreover, the radiation effect would be small because the nanowire is very short. Also, it is well known that silver has a very small emissivity (about 0.03). Therefore, the effect of convection and radiation is negligible in this work. When a constant electrical current is applied through the nanowire, the joule self-heating in the nanowire would induce a temperature rise. The steady-state heat transfer governing equation is as below:

$$k\frac{\partial^2 T}{\partial x^2} + q_0 = 0. \tag{2}$$

Here, $k$ is the thermal conductivity and $q_0$ is the heat generation rate per unit volume. It can be described as $q_0 = I^2 R_1 / (A_c L)$. $I$ is the applied electrical current; $R_1$ is the electrical resistance of the nanowire; $A_c$ is the cross section area of the nanowire ($A_c = \pi d^2 / 4$) and $L$ is the length of the nanowire. The average temperature along the sample is

$$T = \frac{1}{L}\int_{x=0}^{L} T(x)dx = T_0 + \frac{q_0 L^2}{12k}. \qquad (3)$$

The temperature rise is $\Delta T = I^2 R_1 L / (12 k A_c)$. As we can see, the temperature rise is proportional to electrical current's square. When the temperature is higher than 35 K, the electrical resistance of the silver nanowire is proportional to its temperature. The temperature change can be monitored by the electrical resistance variation $\Delta R$:

$$\Delta R = \frac{dR}{dT}\Delta T = \frac{dR}{dT}\cdot\frac{I^2 R_1 L}{12 k A_c}. \qquad (4)$$

It can be seen from Equation (4) that $\Delta R$ is proportional to $I^2$ and the slope is $slope = R_1 L \cdot (dR/dT) / (12 k A_c)$. Then the thermal conductivity of the nanowire can be determined as below:

$$k = \frac{dR}{dT}\cdot\frac{R_1 L}{12 A_c \cdot slope}. \qquad (5)$$

Now we take the thermal conductivity characterization of the silver nanowire at 290 K as an example to demonstrate the measurement process. At first, a series of electrical currents ranging from 0.1 mA to 0.5 mA were applied through the sample and the corresponding voltages were measured by a digit multimeter (Agilent 34401A). Then we plot the relation between the electrical resistance and the current's square as shown in the inset of Fig. 4. After fitting the data linearly, the slope (1.83 Ω/mA$^2$) and intercept (53.15 Ω) can be obtained. The intercept is the electrical resistance ($R_0$) at 290 K. After we measured the electrical resistances at different temperatures, the value of $dR/dT$ (0.1165 Ω/K) can be determined. The diameter and length of the silver nanowire were measured by a scanning electron microscopy (SEM) after the experiment was finished. Finally, the thermal conductivity of the silver nanowire was determined as 191.5 W/m K. In order to improve the measurement accuracy, the electrical resistance rise induced in all the experiments was carefully selected as about 1%. The temperature rises were all controlled within 5 K. The measured temperature dependent thermal conductivity of the silver nanowire is shown and compared with bulk values in Fig. 4.

*4.2 Thermal conductivity of the silver nanowire*

As we can see from Fig. 4, the thermal conductivity of the silver nanowire at 290 K is reduced by 55% from the corresponding bulk's value. Apart from the phonon-electron scattering similar to that in the bulk silver, the extensive structural scatterings, like grain boundary and surface scatterings, also contribute to this reduction. These scatterings limit the electron mean free path and subsequently lead to the reduced thermal conductivity. As the temperature decreases, the thermal conductivity of the silver nanowire behaves totally different from the bulk counterpart. The thermal conductivity of the silver nanowire decreases with decreasing temperature while that of the bulk silver increases with decreasing temperature. Specifically, for bulk silver, the thermal conductivity increases more than ten times when the temperature goes down to 20 K. But for the silver nanowire, the thermal conductivity decreases by 79% when the temperature decreases to 35 K. It is notable that at low temperatures, almost two orders of magnitude reduction in the thermal conductivity was observed for the silver nanowire compared with the bulk silver.

For bulk silver, structural scatterings are rare and phonon scattering dominates the electron transport. When temperature goes down, the short wave phonons are frozen out. The number of excited phonons which involves phonon-electron scattering decreases with decreasing temperature. That is why the thermal conductivity of the bulk silver increases with decreasing temperature. But for the silver nanowire, both structural scatterings and phonon scattering play important roles in the electron transport. When temperature goes down, the phonon scattering diminishes but the structural scatterings still exist and dominate the electron transport. Moreover, the heat capacity of electrons decreases linearly with temperature when the temperature is not too high. That is why the thermal conductivity of the silver nanowire decreases with decreasing temperature. This phenomenon has also been observed in nickel nanowire,[10] gold and platinum nanofilms [13, 28, 29] and alloys [30].

*4.3 Unified thermal resistivity of the silver nanowire*

Here an explanation is provided to the abnormal temperature dependent thermal conductivity of these nanostructures. The thermal conductivity can be written as $\kappa = C_v v_F^2 \tau / 3$. Here $C_v$ is the electron volumetric heat capacity; $v_F$ is the Fermi velocity and $\tau$ is the relaxation time. The volumetric heat capacity of electrons changes linearly with temperature when temperature is not too high ($C_v = \gamma T$). Here $\gamma$ is a constant (0.646 mJ mol$^{-1}$ K$^{-2}$ for silver). The Fermi velocity of silver is $1.39 \times 10^6$ m/s and its electron density is $5.85 \times 10^{28}$ m$^{-3}$.[27] The temperature in the electron heat capacity overshadows the physics of the scattering mechanism behind the variation of thermal conductivity against temperature. Instead of using the traditional thermal resistivity, here we use a unified thermal resistivity $\Theta = T/\kappa$ to explain the thermal conductivity of the nanostructures. The unified thermal resistivity is solely related to the electron relaxation time as below:

$$\Theta = \frac{3}{\gamma v_F^2} \cdot \tau^{-1}. \tag{6}$$

According to the Matthiessen's rule, the unified thermal diffusivity can be separated as two parts: the phonon scattering part and the structural scattering part as below:

$$\Theta = \frac{3}{\gamma v_F^2} \cdot \tau_0^{-1} + \frac{3}{\gamma v_F^2} \cdot \tau_{ph}^{-1} = \Theta_0 + \Theta_{ph}. \tag{7}$$

Getting rid of the effect of temperature on thermal resistivity due to the electron heat capacity, the unified thermal resistivity extracts the effect of temperature on the electron scattering mechanism. The temperature dependent unified thermal resistivity of the silver nanowire and the bulk silver are depicted in Fig. 5. As we can see, the unified thermal resistivity of the silver nanowire ($\Theta_{nanowire}$) and the bulk silver ($\Theta_{bulk}$) shares the same trend when changing against temperature. The two lines are parallel when temperature is not too low. When temperature is above 60 K, the slope of silver nanowire's unified thermal resistivity variation against temperature is $2.57 \times 10^{-3}$ m·K/W and that of the bulk silver is $2.41 \times 10^{-3}$ m·K/W. The slopes are almost the same. This is because the number of excited phonons changes linearly with temperature when temperature is not too low. This confirms that the silver nanowire and the bulk silver share the same phonon-electron scattering mechanism ($\Theta_{ph}$) but have different structural scatterings ($\Theta_0$). The different structural scatterings lead to different residual values ($\Theta_0$) of the unified thermal resistivity. For the bulk silver, the residual unified thermal resistivity is almost zero because the structural imperfection is almost zero in the bulk silver. For the silver nanowire, the residual unified thermal resistivity is large due to grain boundary and surface scatterings which are temperature independent. This residual thermal resistivity can be used to characterize the structure of the silver nanowire because the phonons are frozen out when the temperature approaches zero. The scattering sources are only structural scatterings. According to the residual thermal resistivity (0.9 mK$^2$/W), the relaxation time $\tau_0 = 3/(\gamma v_F^2 \Theta_0)$ is about $2.77 \times 10^{-14}$ s and the characterization length ($l_0 = v_F \tau_0$) is 38.5 nm. This characterization length (electron mean free path limited by structural imperfection) includes the effect of phonon-mediated electron energy transfer across the grain boundaries. It is larger than the real structural size of the crystals in the silver nanowire.

*4.4 Estimation of lattice thermal conductivity of the silver nanowire at low temperatures*

For pure metals, it is well documented that the lattice contribution to the total thermal conductivity is negligible. [31] But for the silver nanowire in this work, the total thermal conductivity at low temperatures is very small. The phonon contribution would be significant. Here we take the case at 36 K as an example to estimate the upper limit of the lattice thermal conductivity. The specific heat of silver at 36 K is 64.65 J/(kg·K) and the density is $10.49 \times 10^3$ kg/m$^3$.[32] The sound speed (2600 m/s) is used to estimate the phonon velocity of silver. [33] At low temperatures, the phonon-phonon scattering mean free path becomes very long due to the decreased phonon density. Moreover, the N-process dominates the phonon-phonon scattering which does not impede heat flux directly and makes little contribution to thermal resistivity. Therefore, at low temperatures the phonon mean free path is limited by the grain boundaries (21 nm). So the upper limit of the thermal conductivity ($\kappa = C_{ph,v} v l / 3$) is calculated as 12.3 W/m·K at 36 K. The phonon mean free path should be smaller than 21 nm because other scatterings like electron-phonon scattering and point defect phonon scattering would also limit the phonon mean free path. The real phonon thermal conductivity should be smaller than 12.3 W/m·K at 36 K. Our measured thermal conductivity is 40.46 W/m·K, so the phonon contribution to the total thermal conductivity is significant, but the electronic thermal conductivity is still dominant at low temperatures.

## 5. Lorenz number's variation with temperature

*5.1 Lorenz number of the silver nanowire*

After the electrical resistivity and thermal conductivity were obtained, it is ready to calculate the Lorenz number. But the measured electrical resistivity is at $T_0$ (the temperature without joule heating) while the measured thermal conductivity is at $T_{\text{ave}}$ (the average temperature during joule heating $T_{\text{ave}} = T_0 + T_1$, $T_1$ is the highest temperature with joule heating). Even though the temperature rise in the measurement process, namely the difference between $T_0$ and $T_1$, is very small (less than 5 K), we cannot calculate the Lorenz number directly. Therefore, linear interpolation was used to obtain the electrical resistivity at $T_{\text{ave}}$. Then the Lorenz number at $T_{\text{ave}}$ was determined as $L_{\text{Lorenz}} = \kappa \rho / T_{\text{ave}}$. Similarly, the Lorenz number at every $T_{\text{ave}}$ can be determined and the temperature dependent Lorenz numbers are shown in Fig. 6.

As we can see from Fig. 6, the Lorenz number at 292 K ($5.2 \times 10^{-8}$ ΩW/K$^2$) is much larger than the Sommerfeld value ($2.44 \times 10^{-8}$ ΩW/K$^2$). Large Lorenz numbers are also observed for nickel nanowire, gold, platinum and Iridium nanfilms. [10, 13-18] This is due to the phonon-assisted electron energy transfer across the grain boundaries. These nanostructures are composed of nanocrystals and there are large number of grain boundaries and surfaces among them. Part of electrons would be reflected back when scattered at the grain boundaries. The reflected electrons can exchange energy with the local phonons. Phonons can transfer across the grain boundaries more readily than electrons. After phonons transfer across the grain boundaries, they can exchange energy with the electrons and phonons in the other side of the grain boundaries. Therefore, when the electrons are reflected back, the charges do not transport through the grain boundaries but part of the electron energy transfers through the grain boundaries. This leads to the greatly reduced electrical conductivity and lesser reduced thermal conductivity.

According to Fig. 6, the Lorenz number of the silver nanowire decreases with decreasing temperature, especially at low temperatures. This is due to the decreasing number of excited phonons and the small angle scattering. On one hand, as the temperature goes down, the number of excited phonons drops. This leads to the decreasing number of phonons which are used to assist to transfer electron energy. Consequently, the Lorenz number at reduced temperature would become smaller than the one at room temperature. On the other hand, as the temperature goes down, only phonons with small wave vector are excited. The wave vector of the phonon population turns gradually towards the lower limit. The electron-phonon scattering would change due to the change of phonon wave vector. Electrons scattering with phonons with large wave vectors are called large angle scattering while electrons scattering with phonons with small wave vectors are called small angle scattering. Large angle scattering impedes the heat and charge transport equally while small angle scattering inhibits the heat transport significantly and leave the charge transport relatively unchanged. [10, 34, 35] At low temperatures, the Lorenz number of the silver nanowire would also decrease due to the extensive small angle scattering.

*5.2 Electrical and thermal electron mean free paths*

The electrical and thermal electron mean free paths shown in the inset of Fig. 6 can be used to interpret the Lorenz number of the silver nanowire. Because free electron model applies to silver, the electrical electron mean free path is calculated from the electrical resistivity ($l = v_F m/(ne^2\rho)$, $v_F$ is the Fermi velocity of silver, $m$ the electron mass, $n$ the electron density of silver and $\rho$ the electrical resistivity). Also, the thermal electron mean free path is calculated from the thermal conductivity [$l = 3\kappa/(\gamma T v_F)$ or $l = 3m\kappa v_F/(\pi^2 n k_B^2 T)$]. As we can see from the inset of Fig. 6, both the electrical and thermal electron mean free path increase with decreasing temperature. This is due to the reduced number of excited phonons and subsequently reduced electron-phonon scattering. The reduced scattering sources extend the electron mean free path. When the temperature approaches absolute zero, all phonons are frozen out. The phonons would not scatter electrons. The only scattering source is structural scatterings, like grain boundary, surface and point defects and these structural scatterings are temperature independent. So the electron mean free path at extremely low temperatures can reflect the structural information of the crystal structure of the silver nanowire. Here the electrical electron mean free path at extremely low temperatures is about 26 nm. This value is consistent with the crystal size (21 nm) in the direction [311] according to the XRD pattern. For the thermal electron mean free path, its value is larger than the electrical mean free path. That is because this value includes the contribution of the phonon-assisted electron energy transfer through the grain boundaries. The difference between the electrical and thermal electron mean free path results in the large Lorenz number. Even though in this work we do not measure the thermal conductivity down to zero, it is predictable that the thermal and electrical mean free path would become the same when temperature approaches the absolute zero. The Lorenz number would become the Sommerfeld value at absolute zero.

## 6. Uncertainty analysis

There are a few factors in the experiment which would affect the accuracy of the measurement results. Here we will have a discussion about them. First, the electrical contact resistance between the silver nanowire and the electrodes is estimated. The Pt pads deposited by EBID are large and good enough to keep good electrical contact. We conducted experiments on silver nanowire without EBID and silver nanowire with silver paste-enhanced contact. In both circumstances, to achieve 1% electrical resistance rise, the applied electrical currents increased after decreasing as the temperature went down. That is because the electrical contact resistance is weakly temperature dependent and the contact resistance dominates the total electrical resistance at low temperatures. The intrinsic electrical resistance needs to rise far more than 1% at low temperatures. That is why large electrical currents are needed at low temperatures. Here the low temperatures means above 25 K because the electrical resistivity becomes weakly temperature dependent when temperature is below 25 K. This would also lead to large applied electrical current to achieve 1% electrical resistance rise. But for the silver nanowire with EBID, the needed electrical currents to achieve 1% resistance rise did not increase at low temperatures. The electrical contact resistance between deposited film and Pt nanowire is also reported negligible in the literature.[11] For the thermal contact resistance after EBID, the Pt-EBID has a contact conductance ($h_{con}$) of 170.5 MW/(m$^2$K) at 293 K.[36] The Pt pad is about 5 μm long for each end and the diameter of the silver nanowire is 227 nm. So the contact area ($A_{con}$) is 1.78 μm$^2$ per end and the thermal contact resistance [$1/(A_{con}h_{con})$] between the silver nanowire and Pt pads is $3.3 \times 10^3$ K/W per end. The two thermal contact resistances are in parallel so the total thermal contact resistance is $1.65 \times 10^3$ K/W. For the silver nanowire, the effective thermal resistance is $\Delta T/q = L/(12\kappa A_c)$. This thermal resistance is defined using the average temperature rise of the sample and the total heat flux through the sample (the joule heat generated by the sample). Here, $L$ and $A_c$ is the silver nanowire length and cross section area. $\kappa$ is the thermal conductivity of the silver nanowire. The thermal resistance of the silver nanowire is $2.9 \times 10^5$ K/W at room temperature. The thermal contact resistance is very small compared with the thermal resistance of the silver nanowire. So the thermal contact resistance is negligible in this work. The thermal contact resistance is also reported negligible in the literature. [11, 36] The length and diameter of the silver nanowire were measured by SEM. The relative errors of the length and diameter measurement are estimated as 1% and 3% respectively. The current error is 0.5% and the voltage error is 0.3%. The relative error of the thermal conductivity and the electrical resistivity are estimated as 7.4% and 4.4% respectively.

## 7. Conclusions

In this work, the thermal and electrical transport properties in an individual silver nanowire were characterized down to 35 K. The results indicated that the thermal and electrical conductivities were significantly reduced compared with their bulk counterparts. The Debye temperature of the silver nanowire (151 K) was found to be reduced by 36% compared with the bulk silver due to phonon softening. The thermal conductivity of the silver nanowire decreased with decreasing temperature while that of the bulk silver increased. To explain the opposite trends of thermal conductivity of the silver nanowire and the bulk silver variation against temperature, a unified thermal resistivity was used to depict the electron scattering mechanism. A large residual unified thermal resistivity (0.9 mK$^2$/W) for the silver nanowire was observed while that of the bulk silver was almost zero. The unified thermal resistivity changed linearly with temperature when the temperature was not too low. This is because the number of the excited phonons decreased linearly with temperature in this temperature range. The unified thermal resistivity of the silver nanowire and the bulk silver shared the same trend, proposing that the silver nanowire and the bulk silver shared the same phonon-electron scattering. Additionally, due to phonon-assisted electron energy transfer across the grain boundaries, the Lorenz number of the silver nanowire ($5.20 \times 10^{-8}$ $\Omega W/K^2$) was found much larger than the bulk counterpart ($2.32 \times 10^{-8}$ $\Omega W/K^2$). Its value decreased with decreasing temperature due to the reduced number of the excited phonons and small angle scattering.

**LIST OF TABLES AND FIGURES**

Figure 1  (a) Schematic diagram of the electrodes and the suspended silver nanowire (top view). (b) SEM picture of the electrodes and the suspended silver nanowire (top view). (c) Schematic diagram of the electrodes and the suspended silver nanowire (side view). (d) SEM picture of the electrodes and the suspended silver nanowire (side view).

Figure 2  XRD pattern of the silver nanowires.

Figure 3  Temperature dependent electrical resistivity of the silver nanowire and the bulk silver.[22] The electrical resistivity of the silver nanowire and the bulk silver are fitted with the Bloch-Grüneisen formula. The obtained Debye temperature of the silver nanowire is 151 K while that of the bulk silver is 235 K. For the inset, it shows the temperature dependent electrical resistivity and electrical resistance of the silver nanowire (the red circles) and they are fitted with the Bloch-Grüneisen formula (the red line).

Figure 4  Temperature dependent thermal conductivity of the silver nanowire and the bulk silver. The inset shows the linear relation between the electrical resistance and the electrical current's square at 290 K during the thermal conductivity measurement of the silver nanowire. The fitting line is $R = 53.15 + 1.833 \times I^2$.

Figure 5  Temperature dependent unified thermal resistivity of the silver nanowire and the bulk silver. When temperature is above 60 K, the slope of silver nanowire's unified thermal resistivity variation against temperature is $2.57 \times 10^{-3}$ m·K/W and that for the bulk silver is $2.41 \times 10^{-3}$ m·K/W.

Figure 6  Temperature dependent Lorenz number of the silver nanowire. The inset shows the temperature dependent thermal and electrical electron mean free paths.

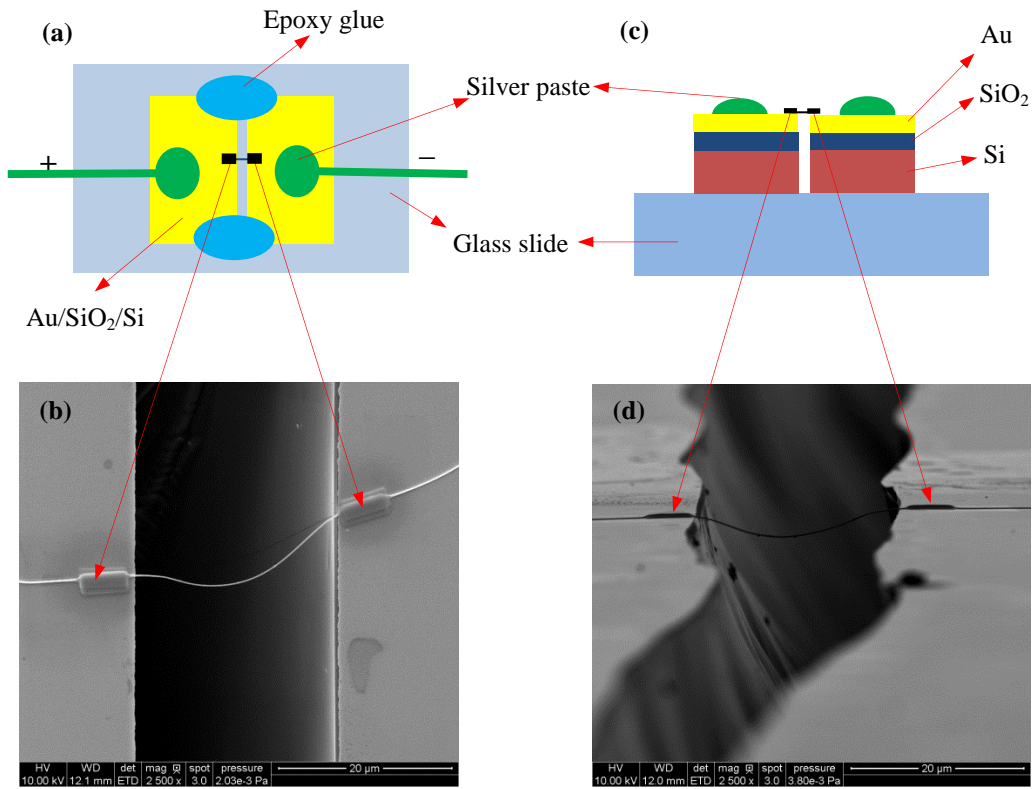

Fig. 1 (a) Schematic diagram of the electrodes and the suspended silver nanowire (top view). (b) SEM picture of the electrodes and the suspended silver nanowire (top view). (c) Schematic diagram of the electrodes and the suspended silver nanowire (side view). (d) SEM picture of the electrodes and the suspended silver nanowire (side view).

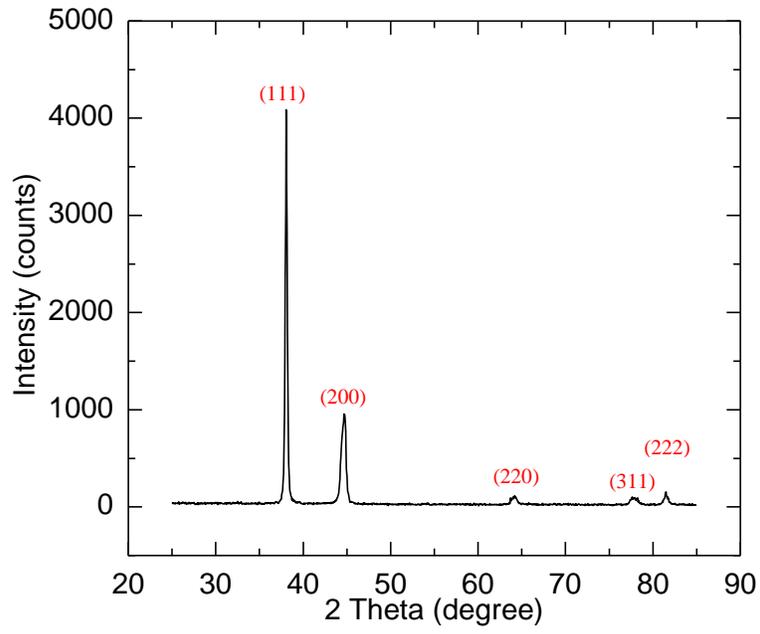

Fig. 2 XRD pattern of the silver nanowires.

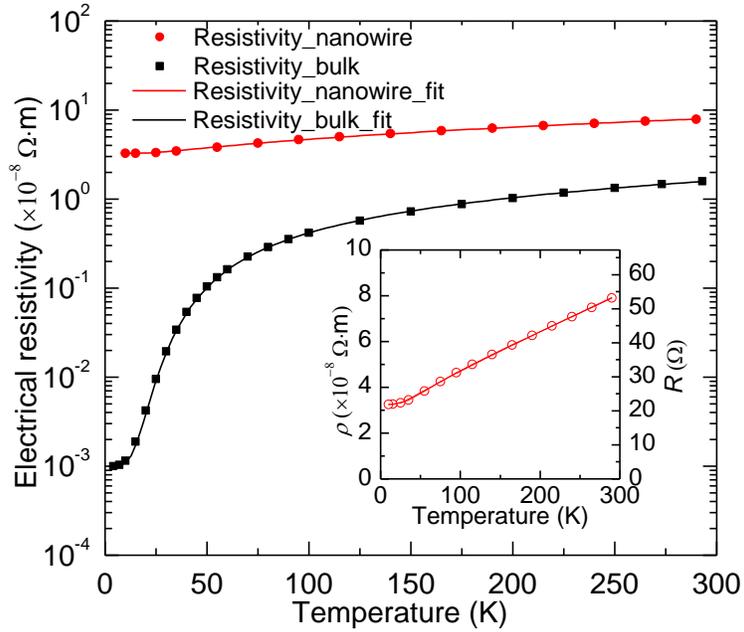

Fig. 3 Temperature dependent electrical resistivity of the silver nanowire and the bulk silver.[22] The electrical resistivity of the silver nanowire and the bulk silver are fitted with the Bloch-Grüneisen formula. The obtained Debye temperature of the silver nanowire is 151 K while that of the bulk silver is 235 K. For the inset, it shows the temperature dependent electrical resistivity and electrical resistance of the silver nanowire (the red circles) and they are fitted with the Bloch-Grüneisen formula (the red line).

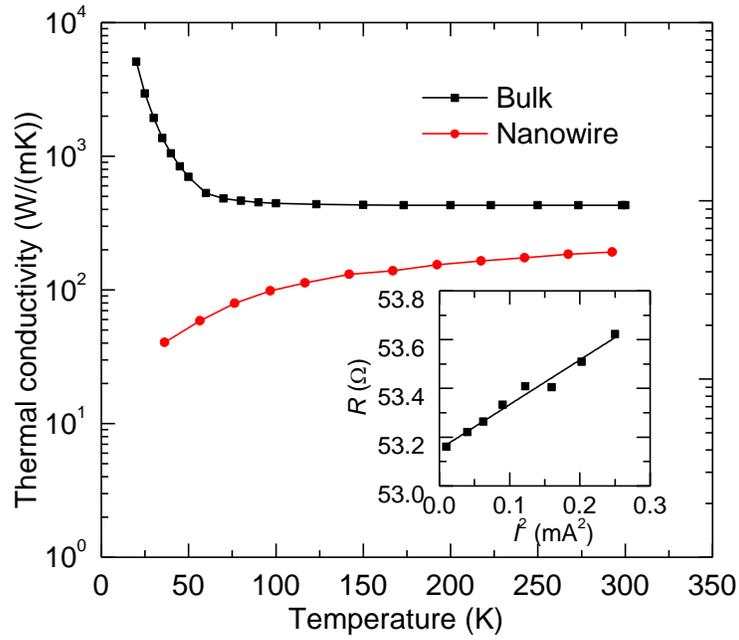

Fig. 4 Temperature dependent thermal conductivity of the silver nanowire and the bulk silver. The inset shows the linear relation between the electrical resistance and the electrical current's square at 290 K during the thermal conductivity measurement of the silver nanowire. The fitting line is $R = 53.15+1.833 \times I^2$.

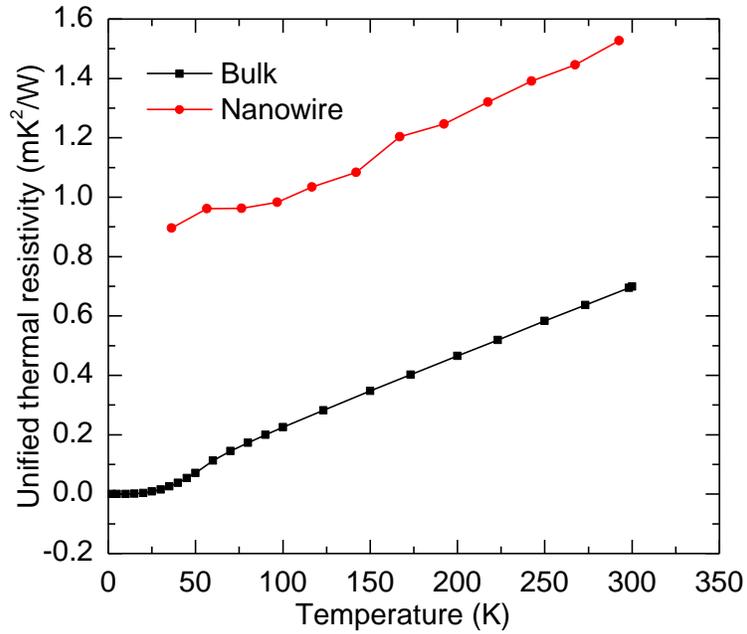

Fig. 5 Temperature dependent unified thermal resistivity of the silver nanowire and the bulk silver. When temperature is above 60 K, the slope of silver nanowire's unified thermal resistivity variation against temperature is $2.57 \times 10^{-3}$ m K/W and that for the bulk silver is $2.41 \times 10^{-3}$ m K/W.

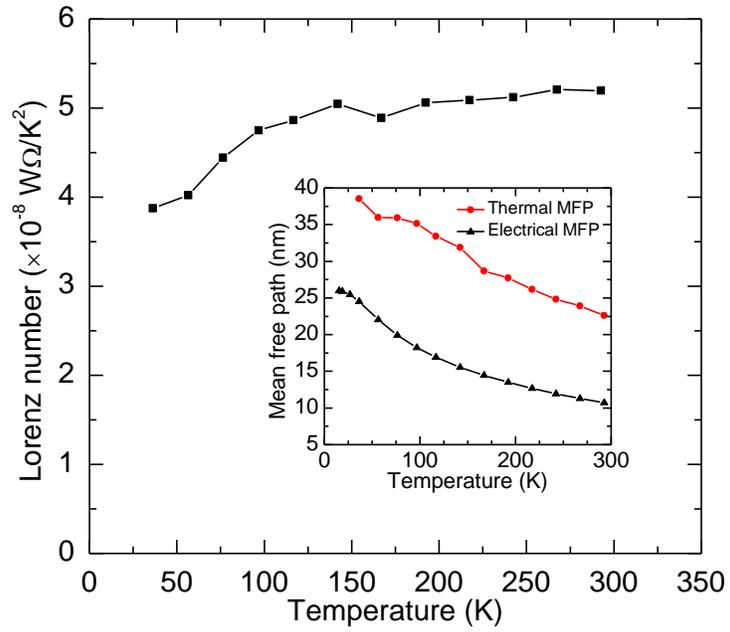

Fig. 6 Temperature dependent Lorenz number of the silver nanowire. The inset shows the temperature dependent thermal and electrical electron mean free paths.